\documentclass[preprint]{revtex4}
\usepackage{amsfonts}
\usepackage{amsmath}
\usepackage{amssymb}
\usepackage{graphicx}

\begin{document}
\title{Scalar leptoquark production at TESLA and CLIC based $e\gamma$ colliders}
\author{O. \c{C}ak{\i}r}
\affiliation{Ankara University, Faculty of Sciences, Department of Physics, 06100,
Tandogan, Ankara, Turkey. }
\author{E. Ateser}
\author{H. Koru}
\affiliation{Gazi University, Faculty of Arts and Sciences, 06500,
Teknikokullar, Ankara, Turkey.}

\begin{abstract}
We study scalar leptoquark production at TESLA and CLIC based $e\gamma$
colliders. Both direct and resolved contributions to the cross section are
examined. We find that the masses of scalar leptoquarks can be probed up to
about 0.9 TeV at TESLA and 2.6 TeV at CLIC.

\end{abstract}
\maketitle

\section{Introduction}

In the standard model (SM) of electroweak (EW) and color (QCD) interactions,
quarks and leptons appear as formally independent components. However, the
observed symmetry between the lepton and quark sectors in the SM could be
interpreted as a hint for common underlying structures. If quarks and leptons
are made of constituents then, at the scale of constituent binding energies,
there should appear new interactions among leptons and quarks. Leptoquarks
(LQs) are exotic particles carrying both lepton number (L) and baryon number
(B), color (anti)-triplet, scalar or vector particles which appear naturally
in various unifying theories beyond the SM. The interactions of LQs with the
known particles are usually described by an effective lagrangian that
satisfies the requirement of baryon and lepton number conservation and
respects the $SU(3)_{C}\times SU(2)_{W}\times U(1)_{Y}$ symmetry of the SM.
There are nine scalar and nine vector leptoquark types according to the BRW
model \cite{1}. The scalar leptoquarks $(S,R)$ can be grouped into singlets
$(S_{0},\widetilde{S}_{0}),$ doublets $(R_{1/2},\widetilde{R}_{1/2})$ and
triplet $(S_{1}).$

The leptoquarks are constrained by different experiments. Direct limits on
leptoquark states are obtained from their production cross sections at
different colliders, while indirect limits are calculated from the bounds on
the leptoquark induced four fermion interactions at low energy experiments.
The mass limits for scalar leptoquarks from single and pair productions
assuming electromagnetic coupling are $M_{LQ}>200$ GeV \cite{2} and
$M_{LQ}>225$ GeV \cite{3}, respectively. Other bounds on the ratio $M_{LQ}/g$
can be obtained from low energy neutral current experiments
(weak charge measurement for Cesium atoms) \cite{4}.

The single production of scalar leptoquarks coupled to $eu$ and $\nu d$
pair in
$e\gamma$ collisions
using only the resolved structure of the photon was
suggested by \cite{5}. The direct single
production of scalar leptoquarks in $e \gamma$ collisions was analyzed in
\cite{6}.

In order to make an analysis with the LQs we make the following
assumptions: The leptoquarks couple to first generation leptons
and quarks, and the couplings $g_{L,R}$ within one generation of
fermions satisfy flavour conservation. The product of couplings
$g_{L}$ and $g_{R}$ vanishes to respect the lepton universality.
One of the scalar leptoquark types gives the dominant contribution
compared with other leptoquark states, and we neglect the
interference between different leptoquark states, i.e. there is no
mixing among LQs. The different leptoquark states within isospin
doublets and triplets are assumed to have the same mass. Under
these assumptions, only the mass and the couplings to right-handed
and/or left-handed leptons, denoted by $g_{R}$ and $g_{L}$, remain
as free parameters.

We study the potential of the TESLA and CLIC based
$e\gamma$ colliders to search for scalar leptoquarks taking into account
both direct and resolved photon processes. We adopt
the
Buchmuller-Ruckl-Wyler (BRW) model \cite{1} which
assumes lepton and baryon number conservation. In the model the interactions
of scalar LQs, having fermion number $F=L+3B$, with fermions can be described
by the effective lagrangian with the dimensionless couplings and
SU(3)$_{C}\times$ SU(2)$_{W}\times$ U(1)$_{Y}$ invariance:
\begin{align}
L_{eff}  &  =L_{F=0}+L_{|F|=2}+L_{\gamma,Z,g}\\
L_{F=0}  &  =g_{1/2L}\overline{u}_{R}l_{L}R_{1/2}+g_{1/2R}
\overline{q}_{L}i\tau_{2}e_{R}R_{1/2}
+\widetilde{g}_{1/2L}\overline{d}_{R}l_{L}\widetilde{R}_{1/2}+\mbox{H.c.}\\
L_{|F|=2}  & =g_{0L}\overline{q}_{L}^{c}i\tau_{2}l_{L}S_{0}+g_{0R}
\overline{u}_{R}^{c}e_{R}S_{0}+\widetilde{g}_{0R}\overline{d}_{R}^{c}
e_{R}\widetilde{S}_{0}+g_{1L}\overline{q}_{L}^{c}i\tau_{2}\overrightarrow{\tau}
l_{L}\cdot\overrightarrow{S}_{1}+\mbox{H.c.}
\end{align}
here the indices of
scalar leptoquarks $S$ or $R$ denote the weak isospin, and an
additional subscript on the couplings indicates the coupled lepton
chirality. A tilde sign is introduced to differentiate between
leptoquarks with different hypercharge. The $l_{L}$ and $q_{L}$
are the left-handed lepton and quark doublets while $e_{R}$ and
$q_{R}$ are the right-handed charged lepton and quark singlets,
respectively. Charged conjugated quark field is defined as $q^{C}%
=C\overline{q}^{T}$ and $\overline{q}^{C}=q^{T}C$.

The gauge interaction of scalar leptoquarks with the EW and QCD
gauge bosons can be described by

\begin{equation}
L_{\gamma,Z,g}=\sum_{\Phi=S,R}\left(  D_{\mu}\Phi\right)  ^{\dagger}\left(
D^{\mu}\Phi\right)  -M_{\Phi}^{2}\Phi^{\dagger}\Phi
\end{equation}
where $\Phi$ is any type of scalar leptoquark, and $M_{\Phi}$ is the mass
of the
scalar leptoquark. The covariant derivative is $D_{\mu}=\partial_{\mu}%
-ig_eQ_SA_{\mu}-ig_eQ_{Z}Z_{\mu}-ig_{s}T^{a}G_{\mu}^{a}$ , here $Q_S$ is
the
charge of scalar leptoquark and $g_{e}$ is the electromagnetic coupling
constant. $Q_{Z}=(I_{3}-Q_S\sin\theta_{W})/\cos\theta_{W}\sin\theta_{W}$.
In the
above equation, $A_{\mu},Z_{\mu}$ and $G_{\mu}$ denote the photon, $Z$-boson
and gluon fields, respectively. $I_{3}$ is the third component of the weak
isospin and $\theta_{W}$ is the Weinberg angle. $T^{a}$ are the Gell-Mann
matrices and $g_{s}$ is the strong coupling constant. From the effective
interaction lagrangian (1) one can deduce quantum numbers of scalar
leptoquarks as
given in Table \ref{table1}.

In this work, a search for singly produced scalar leptoquarks is presented
at
electron-photon colliders. The decay of a heavy LQ into a quark and a charged
lepton leads to final states characterized by an isolated energetic charged
lepton and a hadronic jet, while for decays into a quark and a neutrino, the
final state would have large missing \ energy and a jet. Therefore, under our
assumptions on the couplings, the topologies resulting from the processes
$e\gamma\rightarrow q\overline{q}^{\prime}e$ and $e\gamma\rightarrow
q\overline{q}^{\prime}\nu$ are given in Table \ref{table2}.

\section{Single production of scalar leptoquarks}

Scalar leptoquarks can be produced singly in $e\gamma$ collisions via the
process $e\gamma\rightarrow Sq$ where $S$ is any type of scalar
leptoquarks (S or R).
At $e\gamma$ colliders, with photon beam produced by Compton backscattering,
the maximum $e\gamma$ center of mass energy is about 91\% of available energy.
The projects of TESLA \cite{7} and CLIC \cite{8} colliders will be working
at
$\sqrt{s_{e^{+}e^{-}}}=1$ TeV and $\sqrt{s_{e^{+}e^{-}}}=3$ TeV, respectively.
High energy photon beam can be obtained from the linacs with energies
$\simeq415$ GeV and $\simeq1245$ GeV for TESLA and CLIC, respectively.

\begin{figure}[ptb]
\begin{center}
\includegraphics[
height=2.4898in, width=4.0897in ] {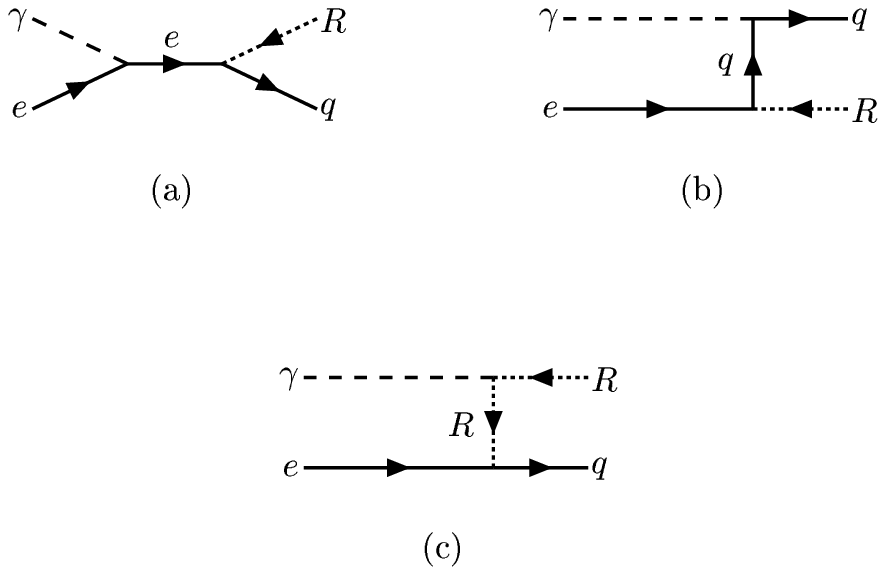}
\end{center}
\caption{Feynman diagrams for $|F|=0$ scalar leptoquarks in
$e\gamma$ collision.} \label{fig1}
\end{figure}

\begin{figure}[ptb]
\begin{center}
\includegraphics[
height=2.4621in, width=4.0067in ] {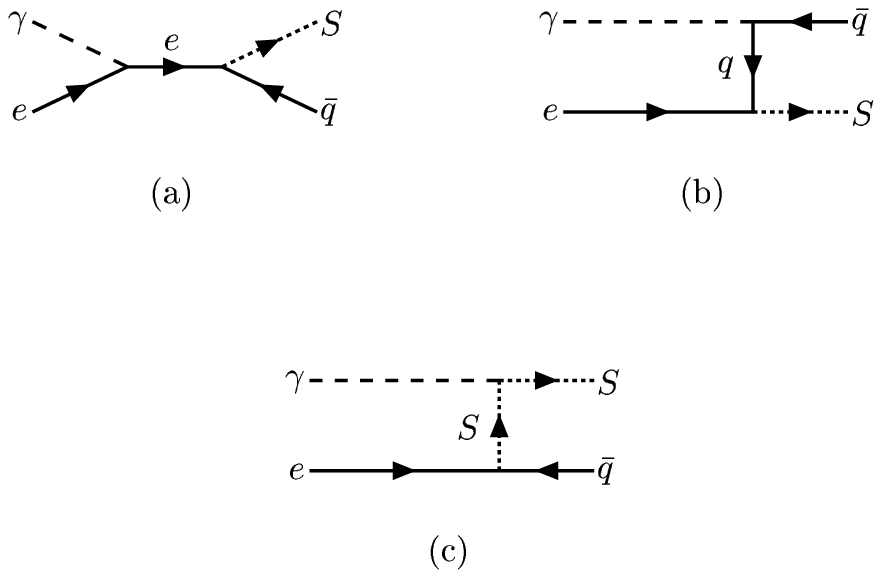}
\end{center}
\caption{Feynman diagrams for $|F|=2$ scalar leptoquarks in
$e\gamma$ collision.} \label{fig2}
\end{figure}

The relevant diagrams for single direct production of scalar
leptoquarks with fermion number $F=0$ ($R$ type) and $|F|=2$ ($S$
type) are shown in Figs. \ref{fig1} and \ref{fig2}. The Feynman
amplitude for the subprocess $e\gamma \longrightarrow Sq$ consists
of $s,u$ and $t$ channels which correspond to electron, quark and
scalar LQ exchanges, respectively.

On the base of the effective lagrangian, the differential cross
section for scalar LQ productions through subprocess
$e\gamma\longrightarrow Sq$ is given by
\begin{align}
\frac{d\widehat{\sigma}}{d\widehat{t}}  &  =\frac{N_{c}g^{2}g_{e}^{2}}%
{16\pi\widehat{s}^{2}}(c_{1}^{2}+c_{2}^{2})\left[  -\frac{Q_{e}^{2}\widehat
{u}}{\widehat{s}}-\frac{Q_{q}^{2}\widehat{s}\widehat{u}}{\left(  \widehat
{u}-m_{q}^{2}\right)  ^{2}}+\frac{Q_{S}^{2}\widehat{t}\left(  \widehat
{t}+M_{LQ}^{2}\right)  }{\left(  \widehat{t}-M_{LQ}^{2}\right)  ^{2}}\right.
\nonumber\\
&  \left.  +\frac{2Q_{e}Q_{q}\left(  \widehat{s}+\widehat{t}\right)  \left(
\widehat{s}-M_{LQ}^{2}\right)  }{\widehat{s}\left(  \widehat{u}-m_{q}%
^{2}\right)  }-\frac{Q_{e}Q_{S}\widehat{t}\left(  \widehat{s}-2M_{LQ}%
^{2}\right)  }{\widehat{s}\left(  \widehat{t}-M_{LQ}^{2}\right)  }-\frac
{Q_{q}Q_{S}\widehat{t}(\widehat{s}+\widehat{t}+M_{LQ}^{2})}{\left(
\widehat{u}-m_{q}^{2}\right)  \left(  \widehat{t}-M_{LQ}^{2}\right)  }\right]
\end{align}
where we use the Lorentz invariant Mandelstam variables
$\widehat{s}=(k_{1}+p_{1})^{2}$, $\widehat{u}=(p_{2}-k_{1})^{2}$
and $\widehat{t}=(k_{2}-k_{1}$)$^{2}$. $c_{1}$ and $c_{2}$ are the
constants equal to $1/2$ and $-(+)1/2$ corresponding to left
(right) couplings, respectively; $g$ is LQ-lepton-quark coupling
constant; $Q_{e},Q_{q}$ and $Q_{S}$ denote electron, quark and LQ
charges, respectively. We denote the
interaction coupling constants of scalar LQs with fermions as $g^{2}%
=4\pi\alpha\kappa$ with $\kappa=1$ and color factor $N_{c}=3$ .
Since cross section varies with $\kappa$ we can simply rescale it
for the various $\kappa$ values.

\begin{figure}[ptb]
\begin{center}
\includegraphics[
height=3.5405in, width=5.0548in ] {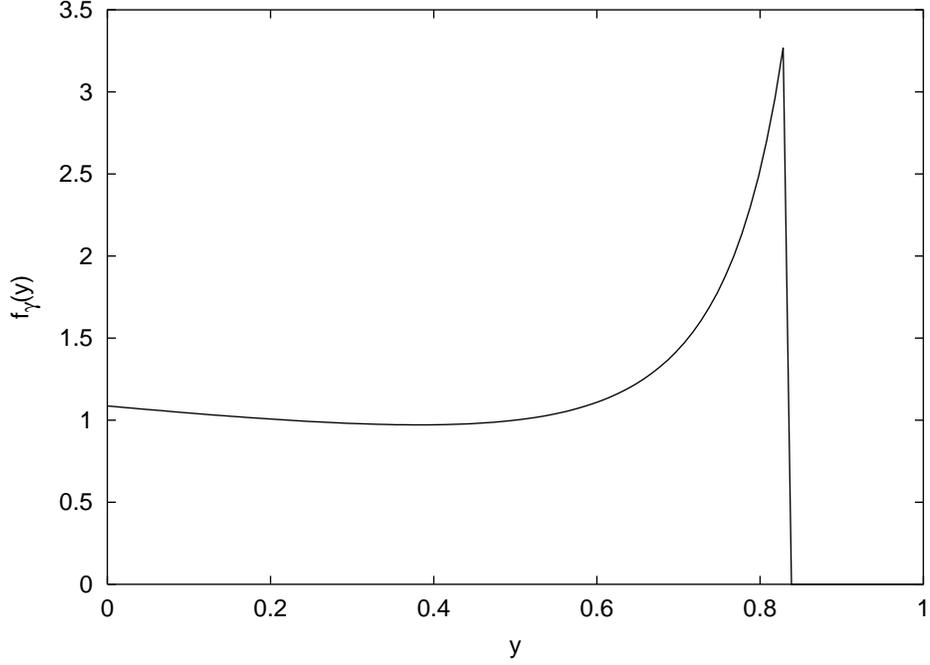}
\end{center}
\caption{Energy spectrum of backscattered photons.} \label{fig3}
\end{figure}

The cross section for the direct production of scalar  leptoquarks via
subprocess $e\gamma
\rightarrow Sq$ is given by%

\begin{equation}
\sigma_{D}=\overset{0.83}{\underset{y_{\min}}{\int}dy}\text{ }f_{y}(y)\text{
}\widehat{\sigma}(y\cdot s)
\end{equation}
where $y_{\min}=M_{LQ}^{2}/s,$ and $f_{\gamma}$ is the
energy spectrum of the Compton backscattered photons from electrons,%

\begin{equation}
f_{\gamma}(y)=\frac{1}{D(\varsigma)}\left[  1-y+\frac{1}{(1-y)}-\frac
{4y}{\varsigma(1-y)}+\frac{4y^{2}}{\varsigma^{2}(1-y)^{2}}\right]
\end{equation}
where%

\begin{equation}
D(\varsigma)=\left(  \frac{1}{2}+\frac{8}{\varsigma}-\frac{1}{2(1+\varsigma
)^{2}}\right)  +\left(  1-\frac{4}{\varsigma}-\frac{8}{\varsigma^{2}}\right)
\ln(1+\varsigma)
\end{equation}
with $\varsigma=4E_{e}\omega_{0}/m_{e}^{2},$ and
$y=E_{\gamma}/E_{e}$ is the ratio of backscattered photon energy
to the initial electron energy. The energy $E_{\gamma}$ of
converted photons is restricted by the condition $y_{\max}=0.83$.
The value $y_{\max}=\varsigma/(\varsigma+1)=0.83$ corresponds to
$\varsigma=4.8$ \cite{9}. Energy spectrum of backscattered photon
is given in Fig. \ref{fig3}. When calculating the production cross
sections for scalar leptoquarks, the divergencies due to the
$u$-channel exchange diagram in Figs. \ref{fig1}b and \ref{fig2}b
are regulated by taking into account the corresponding quark mass
$m_q$. The main contribution to total cross section comes from the
quark exchange diagram in Figs. \ref{fig1}b and \ref{fig2}b. For
this reason, the total cross sections for production of scalar
leptoquarks $R_{1/2}(-5/3),$ $S_{0}(-1/3)$ and $S_{1}(-1/3)$
practically coincide. This is also true for $R_{1/2}(-2/3),$
$\widetilde {R}_{1/2}(-2/3),\widetilde{S}_{0}(-4/3)$ and
$S_{1}(-4/3)$ type scalar leptoquark production.

The scalar leptoquarks can also be produced in $e\gamma$ collisions by the
resolved photon processes, Fig. \ref{fig3}. In order to produce
leptoquarks in
the resonant channel through the quark component of the photon, we study the
following signal for the scalar leptoquark $S$ (or $R$)%

\begin{equation}
e+q_{\gamma}\rightarrow S\rightarrow e+q
\end{equation}
Using the effective lagrangian, the parton level cross section for scalar
leptoquark resonant production can be found as%

\begin{equation}
\widehat{\sigma}(\widehat{s})=\frac{\pi^{2}\kappa\alpha}{2M_{LQ}}\delta
(M_{LQ}-\sqrt{\widehat{s}})
\end{equation}

For CLIC and TESLA based $e\gamma$ colliders, the total cross section for
resolved photon contribution is obtained by convoluting with the backscattered
laser photon distribution and photon structure function. The photon structure
function consists of perturbative pointlike parts and hadronlike parts
$f_{q/\gamma}=f_{q/\gamma}^{PL}+f_{q/\gamma}^{HL}$ \cite{8}$.$ The pointlike
part can be calculated in the leading logaritmic approximation and is given by%

\begin{equation}
f_{q/\gamma}^{PL}(y)=\frac{3\alpha Q_{q}^{2}}{2\pi}\left[  y^{2}%
+(1-y)^{2}\right]  \log\frac{Q^{2}}{\Lambda^{2}}%
\end{equation}
where $Q_{q}$ is the charge of quark content of photon. The cross
section for hadronic contribution
from resolved photon process can be obtained as follows

\begin{equation}
\sigma_{R}=\frac{\pi^{2}\alpha\kappa}{s}\int_{x_{\min}}^{0.83}\frac{dx}%
{x}f_{\gamma/e}(x)\left[  f_{q/\gamma}(z,Q^{2})-f_{q/\gamma}^{PL}%
(z,Q^{2})\right]
\end{equation}
where $z=M_{LQ}^{2}/xs,$ $x_{\min}=M_{LQ}^{2}/s$, and $Q^{2}=M_{LQ}^{2}.$
Since the contribution from the pointlike part of the photon structure
function was already taken into account in the calculation of the direct part
it was subtracted from $f_{q/\gamma}(z,Q^{2})$ in the above formula to
avoid
double counting on the leading logarithmic level. Here, $f_{q/\gamma}$ is a
$Q^{2}$-dependent parton distribution function \cite{8} within the
backscattered photon.

Both schemes give comparable results for the single production cross section.
Therefore, we add their contribution to form the signal. In this
case, the total
cross section for the single production of scalar LQs is

\begin{equation}
\sigma=\sigma_{D}+\sigma_{R}
\end{equation}
here $\sigma_{D}$ and $\sigma_{R}$ denotes the direct and resolved
contribution to the total cross sections. The total cross sections
including both direct and resolved contributions are plotted in
Figs. \ref{fig5}-\ref{fig8} for TESLA and CLIC based $e\gamma$
colliders. In Tables \ref{table3} and \ref{table4}, direct and
resolved contribution to the scalar leptoquark production are
shown for TESLA and CLIC based $e\gamma$ colliders, respectively.
From the figures \ref{fig5}-\ref{fig8} the contribution from
resolved process is dominant for small leptoquark masses. The
distribution function of partons within the photon has the order
$O(\alpha/\alpha_s)$ where $\alpha$ is due to the photon splitting
into $q\bar q$ and $\alpha_s^{-1}$ due to the QCD evolution.
Therefore, the contribution from resolved photon is
$O(\alpha^2/\alpha_s)$ where the hard subprocess in the resolved
photon case has $O(\alpha)$. The direct contribution has
$O(\alpha^2)$. On the other hand, direct contribution is effective
for large masses up to the kinematical limit.

\section{Signals and backgrounds}

When the scalar leptoquarks singly produced at $e\gamma$ colliders
the signal will be double jets and a charged lepton $2j+l$ (S and
R leptoquarks), or double jets plus a neutrino $2j+\not\!p_{T}$ (S
leptoquarks). Since leptoquarks generate a peak in the invariant
($lj$) mass distribution, singly produced leptoquarks are easy to
detect up to the mass values close to the kinematical limit. A
scalar leptoquark decays into a lepton and a quark. The partial
decay width for every decay channel is given by the formula
\begin{equation}
\Gamma=\frac{g^{2}M_{LQ}}{16\pi}
\end{equation}

The leptoquark branchings predicted by the BRW model \cite{1} are given in
Table \ref{table1}. In
the case of $g_{L}=g_{R}$ branchings for
$S_{0}$ can be obtained as $2/3$ for $LQ\rightarrow lq$ and $1/3$ for
$LQ\rightarrow\nu q$ channels. For a given electron-quark, branching ratio
is
defined as $BR(LQ\rightarrow lq)$ and branching ratio to the neutrino-quark is
$BR(LQ\rightarrow\nu q)=1-BR(LQ\rightarrow lq)$ by the definition.

All scalar leptoquark types and signals at the $e\gamma$
collisions are given in Table \ref{table2}. The numbers in
paranthesis denote the leptoquark charge. In order to calculate
the statistical significance $S/\sqrt{B}$ at each mass value of a
scalar leptoquark for each decay channel we need also to calculate
the relevant background cross sections. The Feynman amplitude for
the background subprocess $e\gamma\longrightarrow W^{-}\nu$
consists of $t$ and $s$ channels which correspond to $W^{-}$ and
electron exchanges, respectively. The
differential cross section for this process is given by%

\begin{align}
\frac{d\widehat{\sigma}}{d\widehat{t}}  &  =\frac{-g_{W}^{2}g_{e}^{2}}%
{32\pi\widehat{s}^{3}(\widehat{t}m_{W}-m_{W}^{3})^{2}}\left[  \widehat
{s}\widehat{t}(\widehat{s}^{2}+\widehat{s}\widehat{t}+2\widehat{t}%
^{2})-(3\widehat{s}^{3}+2\widehat{s}^{2}\widehat{t}+5\widehat{s}\widehat
{t}^{2}+\widehat{t}^{3})m_{W}^{2}\right. \nonumber\\
&  \left.  +(\widehat{s}^{2}+5\widehat{t}^{2})m_{W}^{4}+(\widehat{s}%
-5\widehat{t})m_{W}^{6}+m_{W}^{8}\right]
\end{align}
here $\widehat{s}=(p_{e}+p_{\gamma})^{2}$ and $\widehat{t}=(p_{\gamma}%
-p_{W})^{2}$ are the Lorentz invariant Mandelstam variables. The Feynman
amplitude for the background subprocess $e\gamma\longrightarrow Ze$ consists
of $s$ and $u$ channels which both correspond to electron exchanges. The
differential cross section for this subprocess can be written as follows%

\begin{align}
\frac{d\widehat{\sigma}}{d\widehat{t}}  &  =\frac{(c_{V}^{2}+c_{A}^{2}
)g_{e}^{2}g_{z}^{2}}{64\pi\widehat{s}^{3}m_{Z}^{2}(\widehat{s}+\widehat
{t}+m_{e}^{2}-m_{Z}^{2})^{2}}\left
[2(\widehat{s}+\widehat{t})(2\widehat{s}^{2}+2\widehat{s}\widehat
{t}+\widehat{t}^{2}m_{Z}^{2}\right. \nonumber\\
&  \left.  -2(2\widehat{s}+\widehat{t})^{2}m_{Z}^{4}+2(3\widehat
{s}+\widehat{t})m_{Z}^{6}-2m_{Z}^{8})\right]
\end{align}
here $\widehat{s}=(p_{e}+p_{\gamma})^{2}$,
$\widehat{t}=(p_{\gamma}-p_{Z})^{2};$
$c_{V}=-1/2+2\sin^{2}\theta_{W}$ and $c_{A}=-1/2.$

For the background processes $e\gamma\rightarrow W\nu$ and $e\gamma\rightarrow
Ze$ we find the total cross sections $41.20$ $(49.48)$ pb and $2.36$ $(0.49)$
pb at $\sqrt{s_{e^{+}e^{-}}}=1$ $(3)$ TeV, respectively. We multiply
these cross sections
with the branching ratios for corresponding channels. We take the branching
ratios $68.5\%$ and $69.89\%$ for the $W$ boson and $Z$ boson decaying into
hadrons, respectively.

\section{Results and Discussions}

The scalar leptoquarks of any type can be produced with a large cross section
due to direct and resolved processes at $e\gamma$ colliders. The production
cross sections for scalar leptoquarks with the charges $|Q|=5/3$ and $1/3$ are
practically the same, the same situation takes place for the scalar
leptoquarks with the
charges $|Q|=4/3$ and $2/3.$ The contribution to the cross sections from
direct and resolved processes can be compared in Tables \ref{table3} and
\ref{table4}. The resolved contribution is effective relatively at low mass
range. Depending on the center of mass energy this contribution decreases
sharply beyond the leptoquark mass value of about $70\%$ of the collider
energies. The direct contribution for scalar leptoquark with $|Q|=5/3$ is
larger than the scalar leptoquark with $|Q|=4/3.$ This can be explained due to
the quark charge dependence of the cross sections for direct contribution. The
coupling for lepton-quark-leptoquark vertex can be parametrized as $g_{LQ}%
^{2}=4\pi\alpha\kappa$ where $\kappa$ is a parameter.
For smaller values of this parameter the cross section
decreases with $\kappa.$

From the Table \ref{table5}, we find acccesssible upper mass limits of scalar
leptoquarks for TESLA\ based $e\gamma$ collider with the center of mass energy
$\sqrt{s_{e\gamma}^{\max}}\simeq911$ GeV and luminosity $L=10^{5}$ pb$^{-1}$.
The scalar leptoquarks of types $S_{0},S_{1}^{0}$ and $R_{1/2}^{-1/2}$ can be
produced up to mass $M_{LQ}\approx900$ GeV, and $R_{1/2}^{1/2},\widetilde
{S}_{0},S_{1}^{-1},\widetilde{R}_{1/2}^{-1/2}$ up to $M_{LQ}\approx850$ GeV in
the $2j+e$ channel. However, the leptoquarks of type $S_{0},S_{1}^{0}$ can be
produced up to $M_{LQ}\approx850$ GeV and $R_{1/2}^{1/2}$ up to $M_{LQ}%
\approx650$ GeV in the $2j+\not \! p_{T}$ channel. For the CLIC based
$e\gamma$ collider with $\sqrt{s_{e\gamma}^{\max}}\simeq2733$ GeV and
luminosity $L=10^{5}$ pb$^{-1}$, the scalar leptoquarks of type $S_{0}%
,S_{1}^{0},R_{1/2}^{-1/2}$ could be produced up to mass $M_{LQ}\approx2600$
GeV and $\widetilde{R}_{1/2}^{-1/2},\widetilde{S}_{0},S_{1}^{-1}$up to mass
$M_{LQ}\approx2500$ GeV and $R_{1/2}^{1/2}$ up to $M_{LQ}\approx2100$ GeV in
$2j+e$ channel, and scalar leptoquarks of type $R_{1/2}^{1/2}$ up to $700$
GeV, $S_{0}$ up to $900$ GeV and $S_{1}^{0}$ up to $1300$ GeV in the
$2j+\not \! p_{T}$ channel. The statistical significance for these channels
are given in Table \ref{table6}.

As to conclude, the scalar leptoquarks can be produced with a
large numbers at both TESLA and CLIC based $e\gamma$ colliders. We
have analyzed the contributions from direct and resolved photon
processes to the total cross section. We find the latter
contribution is important and can not be ignored especially for
small leptoquark masses. Looking at the final state particles and
their signature in detectors scalar leptoquarks of some types can
be identified.

\newpage

\begin{table}[ptb]
\caption{Quantum numbers of scalar leptoquarks according to BRW
model. The numbers in the parenthesis in the last two columns
denote the values for
$g_{L}=g_{R}.$ }%
\label{table1}
\begin{tabular}
[c]{cccccccc}\hline
Leptoquark & F & I$_{\text{3}}$ & Q$_{\text{em}}$ & Decay & Coupling &
BR($S\rightarrow lq)$ & BR($S\rightarrow\nu q)$\\\hline
$S_{0}$ & 2 & 0 & -1/3 & $%
\begin{tabular}
[c]{c}%
$e_{L}u_{L}$\\
$e_{R}u_{R}$\\
$\nu d_{L}$%
\end{tabular}
\ \ \ \ $ &
\begin{tabular}
[c]{c}%
$g_{0L}$\\
$g_{0R}$\\
$-g_{0L}$%
\end{tabular}
& $\frac{g_{0L}^{2}+g_{0R}^{2}}{2g_{0L}^{2}+g_{0R}^{2}}(\frac{2}{3})$ &
$\frac{g_{0L}^{2}}{2g_{0L}^{2}+g_{0R}^{2}}(\frac{1}{3})$\\\hline
$\widetilde{S}_{0}$ & 2 & 0 & -4/3 & $e_{R}d_{R}$ & $\widetilde{g}_{0R}$ & 1 &
0\\\hline
$S_{1}$ & 0 &
\begin{tabular}
[c]{c}%
1\\
0\\
-1
\end{tabular}
&
\begin{tabular}
[c]{c}%
2/3\\
-1/3\\
-4/3
\end{tabular}
&
\begin{tabular}
[c]{c}%
$\nu u_{L}$\\
$\nu d_{L},e_{L}u_{L}$\\
$e_{L}d_{L}$%
\end{tabular}
&
\begin{tabular}
[c]{c}%
$\sqrt{2}g_{1L}$\\
$-g_{1L}$\\
$-\sqrt{2}g_{1L}$%
\end{tabular}
&
\begin{tabular}
[c]{c}%
0\\
1/2\\
1
\end{tabular}
&
\begin{tabular}
[c]{c}%
1\\
1/2\\
0
\end{tabular}
\\\hline
$R_{1/2}$ & 0 &
\begin{tabular}
[c]{c}%
1/2\\
1/2\\
\\
-1/2\\
-1/2
\end{tabular}
&
\begin{tabular}
[c]{c}%
-2/3\\
-2/3\\
\\
-5/3\\
-5/3
\end{tabular}
&
\begin{tabular}
[c]{c}%
$\nu\overline{u}_{R}$\\
$e_{R}\overline{d}_{L}$\\
\\
$e_{L}\overline{u}_{R}$\\
$e_{R}\overline{u}_{L}$%
\end{tabular}
&
\begin{tabular}
[c]{c}%
$g_{1/2L}$\\
$-g_{1/2R}$\\
\\
$g_{1/2L}$\\
$g_{1/2R}$%
\end{tabular}
&
\begin{tabular}
[c]{c}%
$\frac{g_{1/2R}^{2}}{g_{1/2R}^{2}+g_{1/2L}^{2}}(\frac{1}{2})$\\
\\
1\\
\end{tabular}
&
\begin{tabular}
[c]{c}%
$\frac{g_{1/2L}^{2}}{g_{1/2R}^{2}+g_{1/2L}^{2}}(\frac{1}{2})$\\
\\
0\\
\end{tabular}
\\\hline
$\widetilde{R}_{1/2}$ & 0 &
\begin{tabular}
[c]{c}%
1/2\\
\\
-1/2
\end{tabular}
&
\begin{tabular}
[c]{c}%
1/3\\
\\
-2/3
\end{tabular}
&
\begin{tabular}
[c]{c}%
$\nu\overline{d}_{R}$\\
\\
$e_{L}\overline{d}_{R}$%
\end{tabular}
&
\begin{tabular}
[c]{c}%
$\widetilde{g}_{1/2L}$\\
\\
$\widetilde{g}_{1/2L}$%
\end{tabular}
&
\begin{tabular}
[c]{c}%
0\\
\\
1
\end{tabular}
&
\begin{tabular}
[c]{c}
1\\
\\
0
\end{tabular}
\\\hline
\end{tabular}
\end{table}

\begin{table}[ptb]
\caption{Scalar leptoquark final states, the numbers in the
parenthesis denote the electric charge of scalar leptoquarks.}
\label{table2}
\begin{tabular}
[c]{llllll|lllll}\hline\hline
Initial &  &  &  & Signal &  & Initial &  &  &  & Signal\\\hline
$\gamma e_{L}^{-}$ & $\rightarrow$ & $u_{L}S_{0}(-1/3)$ & $%
\begin{array}
[c]{c}%
\nearrow\\
\searrow
\end{array}
$ & $%
\begin{array}
[c]{c}%
2j+e^{-}\\
2j+\not \!  p_{T}%
\end{array}
$ &  & $\gamma e_{R}^{-}$ & $\rightarrow$ & $u_{R}S_{0}(-1/3)$ &
$\rightarrow$ & $2j+e^{-}$\\\hline
$\gamma e_{L}^{-}$ & $\rightarrow$ & $u_{L}S_{1}(-1/3)$ & $%
\begin{array}
[c]{c}%
\nearrow\\
\searrow
\end{array}
$ & $%
\begin{array}
[c]{c}%
2j+e^{-}\\
2j+\not \!  p_{T}%
\end{array}
$ &  & $\gamma e_{R}^{-}$ & $\rightarrow$ & $d_{R}\widetilde{S}_{0}(-4/3)$
& $\rightarrow$ & $2j+e^{-}$\\\hline
$\gamma e_{L}^{-}$ & $\rightarrow$ & $d_{L}S_{1}(-4/3)$ & $\rightarrow$ &
$2j+e^{-}$ &  & $\gamma e_{R}^{-}$ & $\rightarrow$ & $\overline{d}_{L}
R_{1/2}(-2/3)$ & $\rightarrow$ & $2j+e^{-}$\\\hline
$\gamma e_{L}^{-}$ & $\rightarrow$ & $\overline{u}_{R}R_{1/2}(-5/3)$ &
$\rightarrow$ & $2j+e^{-}$ &  & $\gamma e_{R}^{-}$ & $\rightarrow$ &
$\overline{u}_{L}R_{1/2}(-5/3)$ & $\rightarrow$ & $2j+e^{-}$\\\hline
$\gamma e_{L}^{-}$ & $\rightarrow$ & $\overline{d}_{R}\widetilde{R}_{1/2}
(-2/3)$ & $\rightarrow$ & $2j+e^{-}$ &  &  &  &  &  & \\\hline\hline
\end{tabular}
\end{table}

\begin{table}[ptb]
\caption{The direct and resolved contribution to the cross section for the
scalar leptoquark charges $|Q|=5/3$ and $|Q|=4/3$ at the center of mass energy
$\sqrt{s_{e^{+}e^{-}}}=1$ TeV.}%
\label{table3}
\begin{tabular}
[c]{ccccc}\hline
$\sqrt{s_{e^{+}e^{-}}}=1$ TeV & \multicolumn{2}{c}{$\sigma_{D}($pb$)$} &
\multicolumn{2}{c}{$\sigma_{R}($pb$)$}\\\hline
M$_{\text{LQ}}$(GeV) & $\left\vert Q\right\vert =5/3$ & $\left\vert
Q\right\vert =4/3$ & $\left\vert Q\right\vert =5/3$ & $\left\vert Q\right\vert
=4/3$\\\hline
200 & 2.75 & 0.79 & 1.77 & 1.30\\\hline
300 & 1.86 & 0.53 & 0.61 & 0.42\\\hline
400 & 1.31 & 0.37 & 0.23 & 0.17\\\hline
500 & 0.94 & 0.26 & 0.14 & 7.01$\times$10$^{-2}$\\\hline
600 & 0.69 & 0.19 & 6.39$\times$10$^{-2}$ & 2.46$\times$10$^{-2}$\\\hline
700 & 0.53 & 0.14 & 1.35$\times$10$^{-2}$ & 3.05$\times$10$^{-3}$\\\hline
800 & 0.38 & 9.78$\times$10$^{-2}$ & 0.0 & 0.0\\\hline
900 & 6.63$\times$10$^{-2}$ & 1.66$\times$10$^{-2}$ & 0.0 & 0.0\\\hline
\end{tabular}
\end{table}

\begin{table}[ptb]
\caption{The same as Table \ref{table3}, but for
$\sqrt{s_{e^{+}e^{-}}}=3$ TeV.} \label{table4}
\begin{tabular}
[c]{ccccc}\hline
$\sqrt{s_{e^{+}e^{-}}}=3$ TeV & \multicolumn{2}{c}{$\sigma_{D}$(pb)} &
\multicolumn{2}{c}{$\sigma_{R}$(pb)}\\\hline
M$_{\text{LQ}}$(GeV) & $\left\vert Q\right\vert =5/3$ & $\left\vert
Q\right\vert =4/3$ & $\left\vert Q\right\vert =5/3$ & $\left\vert Q\right\vert
=4/3$\\\hline
300 & 0.53 & 0.16 & 1.27 & 1.09\\\hline
500 & 0.39 & 0.11 & 0.34 & 0.25\\\hline
700 & 0.29 & 8.31$\times$10$^{-2}$ & 0.14 & 9.86$\times$10$^{-2}$\\\hline
900 & 0.23 & 6.37$\times$10$^{-2}$ & 7.43$\times$10$^{-2}$ & 4.76$\times
$10$^{-2}$\\\hline
1100 & 0.18 & 4.97$\times$10$^{-2}$ & 4.32$\times$10$^{-2}$ & 2.54$\times
$10$^{-2}$\\\hline
1300 & 0.14 & 3.92$\times$10$^{-2}$ & 2.75$\times$10$^{-2}$ & 1.41$\times
$10$^{-2}$\\\hline
1500 & 0.11 & 3.12$\times$10$^{-2}$ & 1.82$\times$10$^{-2}$ & 7.78$\times
$10$^{-3}$\\\hline
1700 & 9.34$\times$10$^{-2}$ & 2.52$\times$10$^{-2}$ & 1.18$\times$10$^{-2}$ &
3.94$\times$10$^{-3}$\\\hline
1900 & 7.75$\times$10$^{-2}$ & 2.06$\times$10$^{-2}$ & 6.50$\times$10$^{-3}$ &
1.56$\times$10$^{-3}$\\\hline
2100 & 6.49$\times$10$^{-2}$ & 1.71$\times$10$^{-2}$ & 2.06$\times$10$^{-4}$ &
2.18$\times$10$^{-4}$\\\hline
2300 & 5.34$\times$10$^{-2}$ & 1.38$\times$10$^{-2}$ & 0.0 & 0.0\\\hline
2500 & 3.84$\times$10$^{-2}$ & 9.75$\times$10$^{-3}$ & 0.0 & 0.0\\\hline
2700 & 8.24$\times$10$^{-3}$ & 2.06$\times$10$^{-3}$ & 0.0 & 0.0\\\hline
\end{tabular}
\end{table}

\begin{table}[ptb]
\caption{The number of events and signal significance for $2j+e$ and
$2j+\not \!p_{T}$ channels of scalar leptoquark decays. The lower and
upper indices on scalar leptoquarks $S$ or $R$ denote weak isospin $I$
and $I_3$, respectively.}
\begin{tabular}
[c]{|c|c|c|c|c|c|c|c|c|c|c|}\hline
\multicolumn{1}{|c|}{
\begin{tabular}[c]{c}
$\sqrt{s_{e^{+}e^{-}}}=1$ TeV \\
$L_{int}=10^5$ pb$^{-1}$
\end{tabular}} & \multicolumn{2}{|c|}{
\begin{tabular}[c]{c}
Number of Events\\
$(\sigma_{D}+\sigma_{R})\times L_{int}$
\end{tabular}}
& \multicolumn{5}{|c|}{$\frac{S}{\sqrt{B}}(e\gamma\rightarrow q\overline
{q}e)$} & \multicolumn{3}{|c|}{$\frac{S}{\sqrt{B}}
(e\gamma\rightarrow q\overline{q}^{\prime}\nu)$}\\\hline
M$_{\text{LQ}}$(GeV) & $\left\vert Q\right\vert =5/3$ & $\left\vert
Q\right\vert =4/3$ & $S_{0}$ & $S_{1}^{0}$ & $R_{1/2}^{-1/2}$ & $R_{1/2}
^{1/2}$ & $\widetilde{R}_{1/2}^{-1/2},\widetilde{S}_{0},S_{1}^{-1}$ & $S_{0}$
& $S_{1}^{0}$ & $R_{1/2}^{1/2}$\\\hline
200 & 452407 & 210085 & 704 & 556 & 1113 & 258 & 517 & 89 & 135 & 63\\\hline
300 & 247050 & 94763 & 401 & 304 & 608 & 117 & 233 & 49 & 74 & 28\\\hline
400 & 157382 & 53386 & 255 & 194 & 387 & 66 & 131 & 31 & 47 & 16\\\hline
500 & 107389 & 32904 & 174 & 132 & 264 & 40 & 81 & 21 & 32 & 10\\\hline
600 & 76050 & 21310 & 123 & 94 & 187 & 26 & 52 & 15 & 23 & 6\\\hline
700 & 54541 & 14335 & 88 & 67 & 134 & 18 & 35 & 11 & 16 & 4\\\hline
800 & 38152 & 9782 & 62 & 47 & 94 & 12 & 24 & 7 & 11 & -\\\hline
900 & 6629 & 1660 & 11 & 8 & 16 & 2 & 4 & 1 & 2 & -\\\hline
\end{tabular}
\label{table5}
\end{table}

\begin{table}[ptb]
\caption{The same as Table \ref{table5}, but for CLIC based
$e\gamma$ collider.} \label{table6}
\begin{tabular}
[c]{|c|c|c|c|c|c|c|c|c|c|c|}\hline
\multicolumn{1}{|c|}{
\begin{tabular}[c]{c}
$\sqrt{s_{e^{+}e^{-}}}=3$ TeV \\
$L_{int}=10^5$ pb$^{-1}$
\end{tabular}}
& \multicolumn{2}{|c|}{
\begin{tabular}
[c]{c}
Number of Events\\
$(\sigma_{D}+\sigma_{R})\times L_{int}$
\end{tabular}
} & \multicolumn{5}{|c|}{$\frac{S}{\sqrt{B}}(e\gamma\rightarrow q\overline
{q}e)$} & \multicolumn{3}{|c|}{$\frac{S}{\sqrt{B}}(e\gamma\rightarrow q\overline{q}^{\prime}\nu)$}\\\hline
M$_{\text{LQ}}$(GeV) & $\left\vert Q\right\vert =5/3$ & $\left\vert
Q\right\vert =4/3$ & $S_{0}$ & $S_{1}^{0}$ & $R_{1/2}^{-1/2}$ & $R_{1/2}
^{1/2}$ & $\widetilde{R}_{1/2}^{-1/2},\widetilde{S}_{0},S_{1}^{-1}$ & $S_{0}$
& $S_{1}^{0}$ & $R_{1/2}^{1/2}$\\\hline
300 & 181523 & 124657 & 648 & 491 & 981 & 337 & 674 & 33 & 49 & 34\\\hline
500 & 73182 & 35515 & 261 & 198 & 396 & 99 & 197 & 13 & 20 & 10\\\hline
700 & 43722 & 18168 & 156 & 118 & 236 & 49 & 98 & 8 & 12 & 5\\\hline
900 & 30084 & 11135 & 107 & 81 & 163 & 30 & 60 & 5 & 8 & 3\\\hline
1100 & 22141 & 7513 & 79 & 60 & 120 & 20 & 41 & 4 & 6 & -\\\hline
1300 & 16929 & 5335 & 60 & 46 & 92 & 14 & 29 & - & 5 & -\\\hline
1500 & 13251 & 3906 & 47 & 36 & 72 & 11 & 21 & - & 4 & -\\\hline
1700 & 10520 & 2922 & 38 & 28 & 57 & 8 & 16 & - & - & -\\\hline
1900 & 8406 & 2225 & 30 & 23 & 45 & 6 & 12 & - & - & -\\\hline
2100 & 6705 & 1728 & 24 & 18 & 36 & 5 & 9 & - & - & -\\\hline
2300 & 5340 & 1377 & 19 & 14 & 29 & 4 & 7 & - & - & -\\\hline
2500 & 3842 & 975 & 14 & 10 & 21 & - & 5 & - & - & -\\\hline
2700 & 824 & 206 & 3 & 2 & 4 & - & 1 & - & - & -\\\hline
\end{tabular}
\end{table}

\newpage

\begin{figure}[ptb]
\begin{center}
\includegraphics[
height=1.2246in, width=3.3062in ] {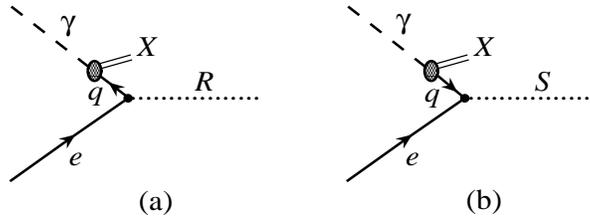}
\end{center}
\caption{Resolved process for single production of scalar
leptoquarks. a) correspond to $F=0$, and b) for $\vert F\vert=2$
type.} \label{fig4}
\end{figure}

\begin{figure}[ptb]
\begin{center}
\includegraphics[
height=3.5405in, width=5.0548in ] {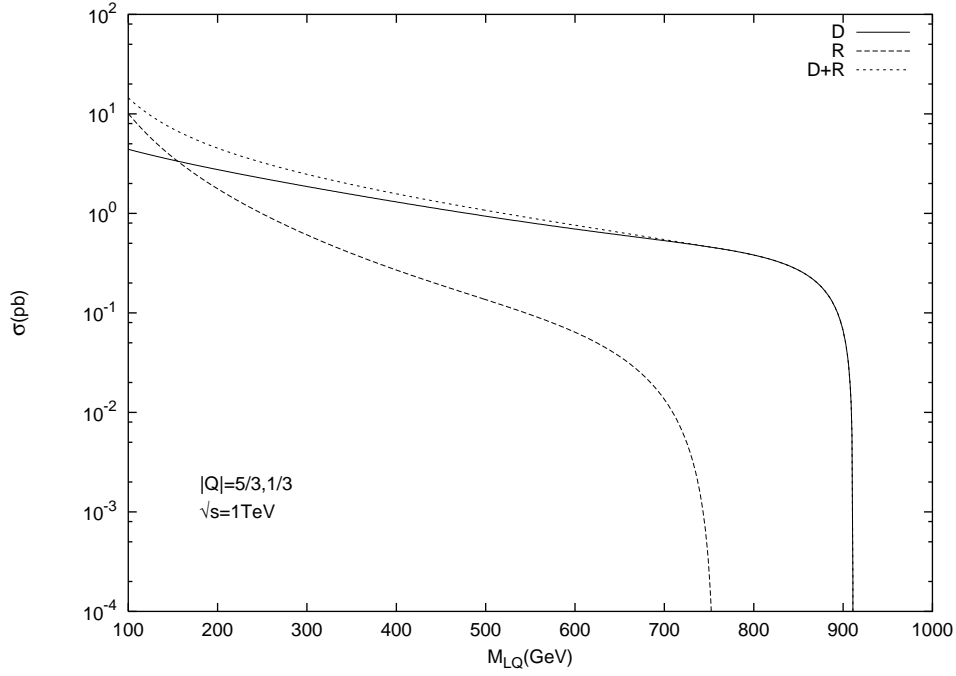}
\end{center}
\caption{The direct (D) and resolved (R) contributions to the
cross section depending on scalar leptoquark mass M$_{LQ}$ with
charge $|Q|=5/3(1/3)$ and $\sqrt{s_{e^{+}e^{-}}}=1$ TeV. }
\label{fig5}
\end{figure}

\begin{figure}[ptb]
\begin{center}
\includegraphics[
height=3.5405in, width=5.0548in ] {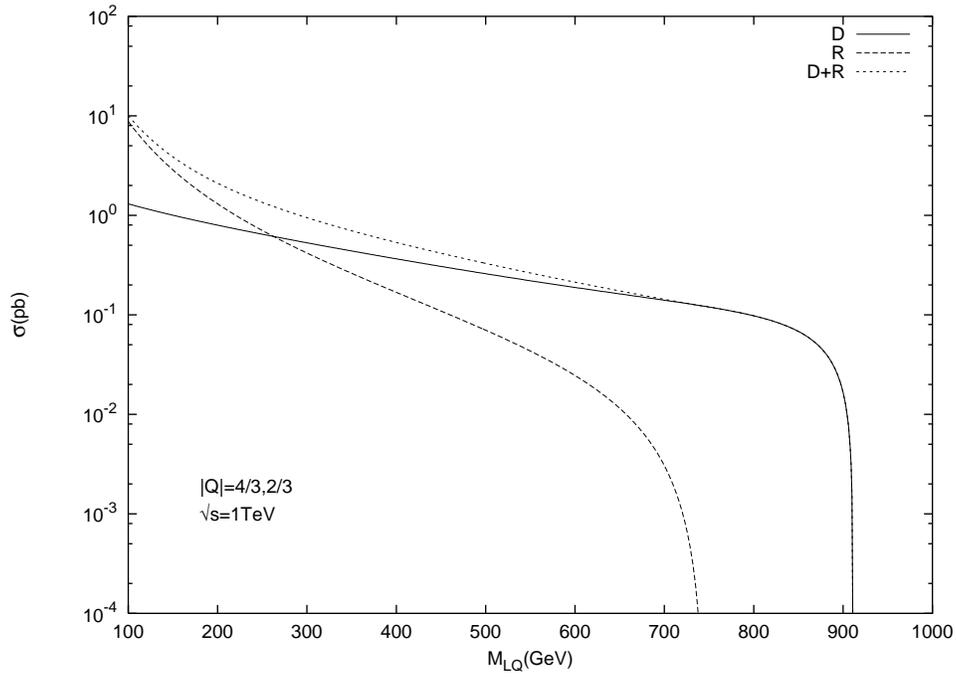}
\end{center}
\caption{The same as Fig. \ref{fig5}, but for the charge
$|Q|=4/3(2/3)$. } \label{fig6}
\end{figure}

\begin{figure}[ptb]
\begin{center}
\includegraphics[
height=3.5405in, width=5.0548in ] {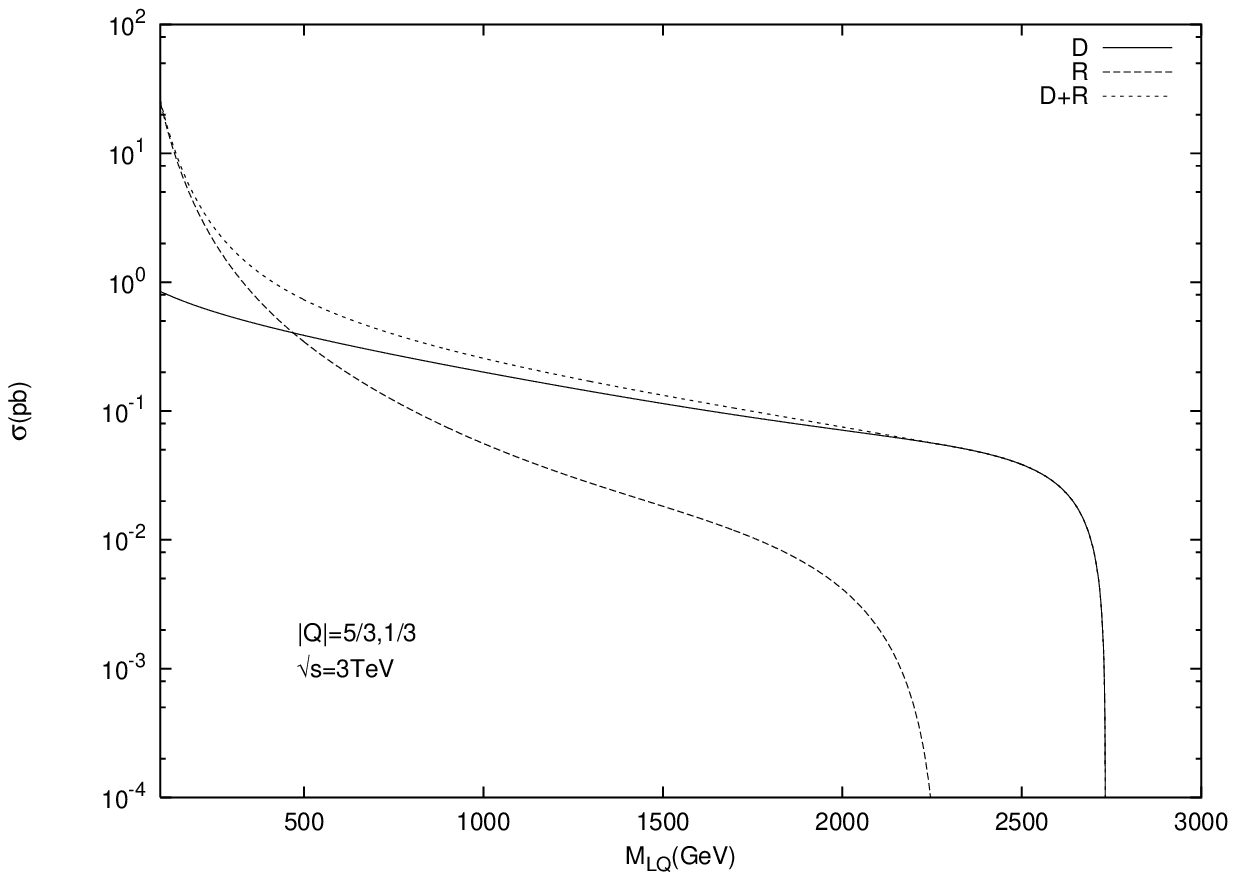}
\end{center}
\caption{The same as Fig. \ref{fig5}, but for the center of mass
energy $\sqrt{s_{e^{+}e^{-}}}=3$ TeV. } \label{fig7}
\end{figure}

\begin{figure}[ptb]
\begin{center}
\includegraphics[
height=3.5405in, width=5.0548in ] {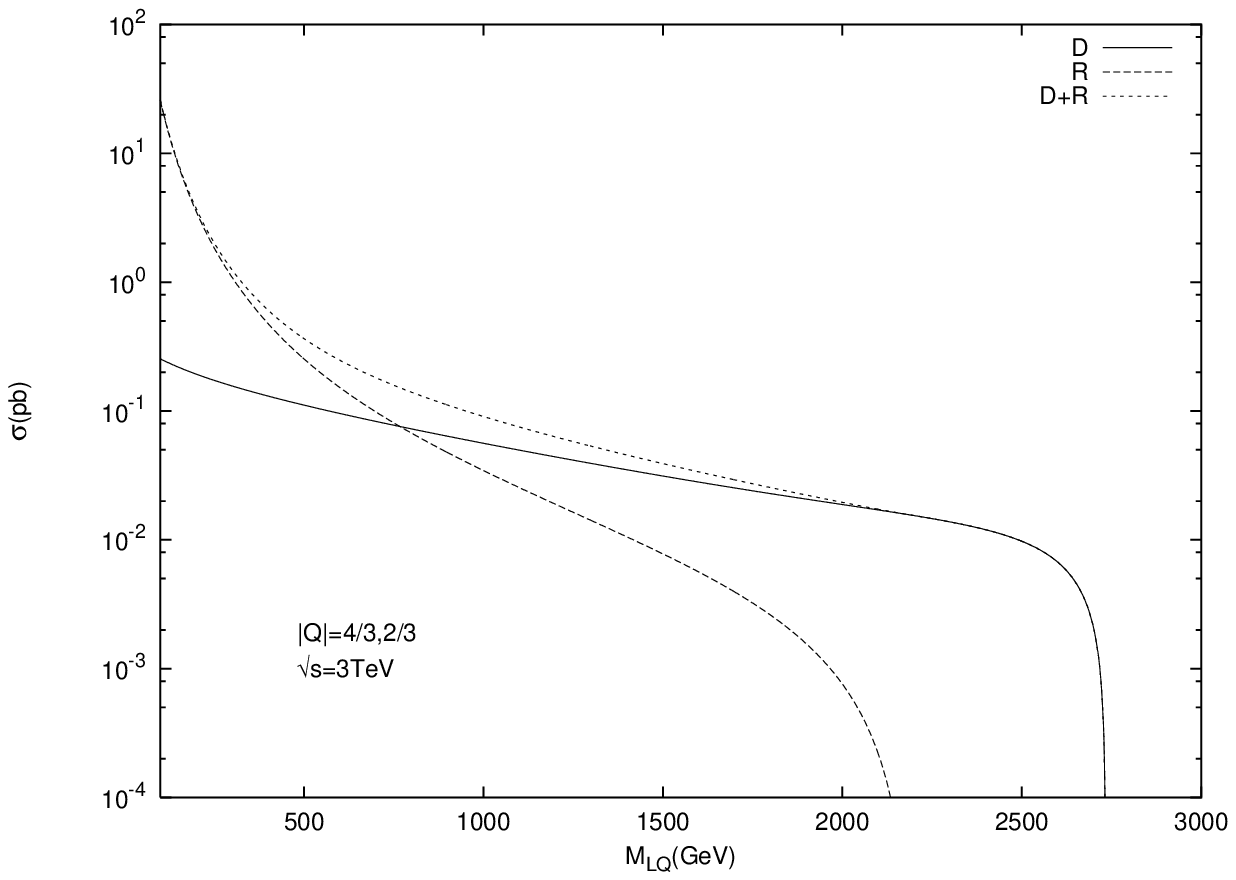}
\end{center}
\caption{The same as Fig. \ref{fig6}, but for the center of mass
energy $\sqrt{s_{e^{+}e^{-}}}=3$ TeV. } \label{fig8}
\end{figure}

\end{document}